\documentclass[graphicx]{iopart}
\usepackage{iopams,setstack}
\usepackage{epsfig}

\begin{document}

\title[Laser-induced nonsequential double ionization in diatomic molecules]{%
Laser-induced nonsequential double ionization in diatomic molecules: one and
two-center rescattering scenarios}
\author{C. Figueira de Morisson Faria}
\date{\today}

\begin{abstract}
We investigate laser-induced nonsequential double ionization from aligned
diatomic molecules, using the strong-field approximation in its length and
velocity gauge formulations. Throughout, we consider that the first electron
dislodges the second by electron-impact ionization. Employing modified
saddle-point equations, we single out the contributions of different
scattering scenarios to the maxima and minima observed in the differential
electron momentum distributions. We show that the quantum interference
between the electron orbits starting and ending at a specific center $C_{j}$%
, and those starting at $C_{j}$ and ending at a different center $C_{\nu }$
leads to the same maxima as minima as if all possible scenarios are taken.
There exist, however, quantitative differences as far as the gauge choice is
concerned. Indeed, while the velocity-gauge distributions obtained employing
only the above-mentioned processes are practically identical to the overall
distributions, their length-gauge counterparts exhibit an asymmetry in the
positive and negative momentum regions. This asymmetry is due to additional
potential-energy shifts which are only present in the length-gauge
formulation, and which, depending on the center, sink or increase the
potential barrier through which the first electron tunnels. In contrast, the
interference between topologically similar scenarios leads at most to
patterns whose positions, in momentum space, do not agree with the overall
interference condition, neither in the length nor in the velocity gauge.
\end{abstract}

\pacs{32.80.Rm, 33.80.Rv}

\address{Department of Physics and Astronomy, University College
London,\\ Gower Street, London WC1E 6BT, United
Kingdom}\ead{c.faria@ucl.ac.uk}

\section{Introduction}

Molecules in strong laser fields $(I\sim 10^{14}-10^{15}\mathrm{W/cm}^{2})$
have been the object of intensive scrutiny in the past few years. In
particular, high-order harmonic generation (HHG) and above-threshold
ionization (ATI) have been employed to extract information about the
structure of such systems with subfemtosecond precision \cite{attomol}. This
is made possible by the physical mechanisms behind such phenomena, namely
the rescattering or recombination of the active electron with its parent
molecule \cite{tstep}. These processes take place within a fraction of a
laser cycle. For a typical Titanium-Sapphire laser field, whose cycle is
roughly $T\sim 2.7fs$, this corresponds to hundreds of attoseconds \cite%
{Scrinzi2006}. In particular, the configuration of atoms with which the
electron rescatters or recombines leads to patterns which are characteristic
of the molecule. These patterns may be either due to the quantum
interference of photoelectron or harmonic emission in spatially separated
centers \cite{doubleslit}, or of the rescattering or recombination scenarios
involving more than one center \cite%
{KBK98,Usach2006,PRACL2006,HBF2007,F2007,F2008,DM2008}.

Specifically for ATI and HHG in diatomic molecules, there exist analytic
expressions which give the approximate energy positions of the interference
minima and maxima due to the above-mentioned spatial separation \cite%
{doubleslit}. This allows a physical interpretation in terms of a
microscopic double-slit experiment. In particular in the framework of
semi-analytic, S-Matrix approaches, such as the strong-field approximation
(SFA), the HHG or ATI transition amplitudes are written as multiple
integrals, with slowly-varying prefactors and a semiclassical action \cite%
{SFAold}. In this case, the double-slit interference appears as a prefactor,
which depends on the symmetry of the highest occupied molecular orbital and
on the internuclear distance \cite%
{KBK98,Usach2006,PRACL2006,HBF2007,F2007,DM2008,Usachenko,moreCL,MBBF00,Madsen,DM2006,Kansas}%
.

There are, however, comparatively fewer studies of the influence of
two-center scenarios, in which an electron is released in a center $C_{j}$
in the molecule and rescatters or recombines with a different center $C_{\nu
},$ $j\neq \nu $ \cite{KBK98,Usach2006,PRACL2006,HBF2007,F2007,F2008}$.$ In
these references, the two-center processes have also been treated and
discussed in quite different ways. For instance, in \cite%
{PRACL2006,F2007,F2008}, the high-order harmonic double-slit prefactor has
been exponentialized and incorporated in the action, while in high-order ATI
the two-center processes led to transition amplitudes, which could not be
grouped as to provide a common prefactor \cite{HBF2007}.

In previous work, we have shown that well-defined interference fringes may
be also present in laser-induced nonsequential double ionization (NSDI) of
diatomic molecules \cite{FSLY2008}, for a small range of alignment angles.
This is a consequence of the fact that NSDI is also described as a
laser-induced rescattering process. More specifically, the first electron,
freed at a time $t^{\prime }$, is accelerated by the external laser field
and, at a subsequent time $t$, collides inelastically with its parent
molecule$.$ In this collision, it gives part of the kinetic energy it
acquired from the field to a second electron, which is then released.
Quantum mechanically, transition amplitudes corresponding to NSDI at
different centers in the molecule are expected to interfere. Even though, up
to the present date, there is no experimental evidence of such fringes,
recently, it has been observed that the shapes and the peaks of the NSDI
differential electron momentum distributions in molecules depend on the
symmetry of the highest occupied molecular orbital \cite{NSDIsymm}, and on
the molecular alignment angle with respect to the laser-field polarization
\cite{NSDIalign}.

In \cite{FSLY2008}, we worked mainly within the SFA, and assumed that the
second electron was dislodged by the simplest possible physical mechanism:
electron-impact ionization. We have shown that, in this case, the two-center
interference leads to minima and maxima parallel to the anti-diagonal $%
p_{(1)\parallel }+p_{(2)\parallel }$ in the plane of the momentum components
$(p_{(1)\parallel },p_{(2)\parallel })$ parallel to the laser-field
polarization. For parallel-aligned molecules, these fringes are sharpest. As
the alignment angle increases, the contributions from the perpendicular
momentum components start to blur these fringes, until, for perpendicular
alignment, all structure is washed out. In such investigations, we
incorporated the structure of the molecule in the prefactor, and kept the
same action as in the single-atom case.

A legitimate question is, however, how the different rescattering scenarios,
involving one or two centers, contribute to the above-mentioned patterns in
NSDI. In the following, we will address this issue, incorporating the
prefactors derived in \cite{FSLY2008} in the semiclassical action. A closely
related procedure has been followed in \cite{PRACL2006,F2007}, for
high-order harmonic generation. In order to facilitate this assessment and
make the two-center interference fringes as clear as possible, we strip the
problem to its bare bones using a series of assumptions. Firstly, unless
otherwise stated, we consider very simplified highest occupied molecular
orbitals, for which analytic interference conditions have been derived \cite%
{FSLY2008}. Furthermore, we choose the interaction through which the second
electron is dislodged as a contact-type interaction at the positions of the
ions. As it will be discussed subsequently (c.f. Sec. \ref{results}), this
type of interaction eliminates any bias in the electron momentum
distributions, which may be detrimental for their interpretation\footnote{%
Apart from that, in the single-atom case and within the SFA framework, this
type of interaction led to a better agreement with the existing experiments
than more realistic choices, such as a Coulomb-type interaction. For a
detailed discussion, see Refs. \cite{FSLB2004,FL2005}.}. Finally,
throughout, we will consider the condition for which the interference
patterns are most pronounced, i.e., parallel alignment.

This paper is organized as follows. In Sec. \ref{transampl}, we give the
expression for the transition amplitude related to the process in which the
second electron is freed by electron-impact ionization, within the SFA. We
also briefly recall the two-center prefactors employed in \cite{FSLY2008},
and the expression derived in Ref. \cite{FSLY2008} for the two-center NSDI
interference conditions (Sec. \ref{prefactors}). Subsequently, in Sec. \ref%
{saddmodified}, we provide the expressions for the modified saddle-point
equations, in which different scattering processes are taken into account.
In Sec. \ref{results}, these equations are employed to compute differential
momentum distributions, which are analyzed in detail. Finally, in Sec. \ref%
{concl}, we summarize the paper and provide its main conclusions.

\section{Transition amplitudes}

\label{transampl}

\subsection{General expressions}

The simplest process responsible for laser-induced nonsequential double
ionization is electron-impact ionization. Within the Strong-Field
Approximation, the corresponding transition amplitude is given by
\begin{equation}
\hspace*{-1.2cm}M(\mathbf{p}_{(1)},\mathbf{p}_{(2)})=\int_{-\infty }^{\infty
}dt\int_{-\infty }^{t}dt^{\prime }\int d^{3}kV_{\{\mathbf{p}%
_{(n)}\}k}V_{k0}\exp [iS(\{\mathbf{p}_{(n)}\},\mathbf{k},t,t^{\prime })],
\label{Mp}
\end{equation}%
with
\begin{eqnarray}
\hspace*{-1.2cm}S(\{\mathbf{p}_{(n)}\},\mathbf{k},t,t^{\prime })
&=&-\sum_{n=1}^{2}\int_{t}^{\infty }\hspace{-0.1cm}\frac{[\mathbf{p}_{(n)}+%
\mathbf{A}(\tau )]^{2}}{2}d\tau -\int_{t^{\prime }}^{t}\hspace{-0.1cm}\frac{[%
\mathbf{k}+\mathbf{A}(\tau )]^{2}}{2}d\tau  \label{singlecS} \\
&&-E_{0(2)}t-E_{0(1)}t^{\prime }  \nonumber
\end{eqnarray}%
and the prefactors%
\begin{equation}
V_{\mathbf{k}0}=\left\langle \mathbf{\tilde{k}}(t^{\prime })\right\vert
V\left\vert \phi _{0}^{(1)}\right\rangle  \label{Vk0}
\end{equation}%
and%
\begin{equation}
V_{\{\mathbf{p}_{(n)}\}\mathbf{k}}=\left\langle \mathbf{\tilde{p}}%
_{(1)}\left( t\right) ,\mathbf{\tilde{p}}_{(2)}\left( t\right) \right\vert
V_{12}\left\vert \mathbf{\tilde{k}}(t),\phi _{0}^{(2)}\right\rangle ,
\label{Vpnk}
\end{equation}%
where the indices in parentheses are related to each electron involved and
the term $\{\mathbf{p}_{(n)}\}$ in brackets is related to \textit{both}
electron momenta. Eq.~(\ref{Mp}) describes the process in which an electron,
initially in a bound state $\left\vert \phi _{0}^{(1)}\right\rangle ,$ is
freed by tunneling ionization at a time $t^{\prime }$ into a Volkov state $%
\left\vert \mathbf{\tilde{k}}(t)\right\rangle $. Subsequently, this electron
propagates in the continuum from $t^{\prime }$ to a later time $t.$ At this
time, the electron collides inelastically with its parent molecule, and a
second electron, which is bound in $\left\vert \phi _{0}^{(2)}\right\rangle $%
, is then released through the interaction $V_{12}.$ Thereafter, both
electrons are in Volkov states, and their final momenta are $\mathbf{p}%
_{(n)}(n=1,2).$ In the above-stated equations, $E_{0(n)}(n=1,2)$ give the
first and second ionization potentials, and $V$ the atomic potential. The
form factors (\ref{Vk0}) and (\ref{Vpnk}) contain all the information about
the atomic potential, and the interaction by which the second electron is
dislodged (see \cite{FSLB2004} for details). One should note that the
transition amplitude cannot be factorized into one-electron transition
amplitudes. Physically, this implies a strongly correlated two-electron
process.

Another noteworthy issue is that, within the SFA, $V_{\mathbf{k}0}$ and $%
V_{\{\mathbf{p}_{(n)}\}\mathbf{k}}$ are gauge dependent. In fact, in the
length gauge $\mathbf{\tilde{p}}_{(n)}\left( \tau \right) =\mathbf{p}_{(n)}+%
\mathbf{A}(\tau )$ and $\mathbf{\tilde{k}}(\tau )=\mathbf{k}+\mathbf{A}(\tau
)(\tau =t,t^{\prime }),$ while in the velocity gauge $\mathbf{\tilde{p}}%
_{(n)}\left( \tau \right) =\mathbf{p}_{(n)}$ and $\mathbf{\tilde{k}}(\tau )=%
\mathbf{k}.$ A similar gauge dependence has also been reported for
high-order harmonic generation \cite{PRACL2006,F2007} and above-threshold
ionization \cite{DM2006,DM2008}. \ For a more general discussion see \cite%
{FKS96}.

\subsection{Two-center prefactors}

\label{prefactors} In the specific case of diatomic molecules, we will
consider the same simplified model as in \cite{FSLY2008}, namely frozen
nuclei, the linear combination of atomic orbitals (LCAO) approximation, and
homonuclear molecules. Under these assumptions, the molecular bound-state
wave function for each electron reads
\begin{equation}
\psi _{0}^{(n)}(\mathbf{r}_{(n)})=C_{\psi }\left[ \phi _{0}^{(n)}(\mathbf{r}%
_{(n)}-\mathbf{R}/2)+\epsilon \phi _{0}^{(n)}(\mathbf{r}_{(n)}+\mathbf{R}/2)%
\right] ,
\end{equation}%
where $n=1,2,$ $\epsilon =\pm 1$, and $C_{\psi }=1/\sqrt{2(1+\epsilon S(%
\mathbf{R})}$, with
\begin{equation}
S(\mathbf{R})=\int \left[ \phi _{0}^{(n)}(\mathbf{r}_{(n)}-\mathbf{R}/2)%
\right] ^{\ast }\phi _{0}^{(n)}(\mathbf{r}_{(n)}+\mathbf{R}/2)d^{3}r.
\end{equation}%
\ The positive and negative signs for $\epsilon $ correspond to bonding and
antibonding orbitals, respectively.

The molecular binding potential, as seen by each electron, is written as
\begin{equation}
V(\mathbf{r}_{(n)})=V_{0}(\mathbf{r}_{(n)}-\mathbf{R}/2)+V_{0}(\mathbf{r}%
_{(n)}+\mathbf{R}/2),
\end{equation}%
where $V_{0}$ corresponds to the binding potential of each center in the
molecule, which, at this stage, are kept general. The above-stated
assumptions yield
\begin{equation}
V_{\mathbf{k}0}^{(b)}=-\frac{2C_{\psi }}{(2\pi )^{3/2}}\cos [\mathbf{\tilde{k%
}}(t^{\prime })\cdot \mathbf{R}/2]\mathcal{I}(\mathbf{\tilde{k}}(t^{\prime
}))  \label{Vk0bond}
\end{equation}%
or%
\begin{equation}
V_{\mathbf{k}0}^{(a)}=-\frac{2iC_{\psi }}{(2\pi )^{3/2}}\sin [\mathbf{\tilde{%
k}}(t^{\prime })\cdot \mathbf{R}/2]\mathcal{I}(\mathbf{\tilde{k}}(t^{\prime
})),  \label{Vk0anti}
\end{equation}%
for the bonding and antibonding cases, respectively, with%
\begin{equation}
\mathcal{I}(\mathbf{\tilde{k}}(t^{\prime }))=\int d^{3}r_{(1)}\exp [i\mathbf{%
\tilde{k}}(t^{\prime })\cdot \mathbf{r}_{(1)}]V_{0}(\mathbf{r}_{(1)})\phi
_{0}^{(1)}(\mathbf{r}_{(1)}).  \label{ik}
\end{equation}%
In the above-stated equations, we have neglected the integrals for which the
binding potential and the bound-state wave function are localized at
different centers in the molecule, due to the fact that they are very small
for the parameter range of interest.

If the electron-electron interaction depends only on the distance between
the two electrons, i.e., $V_{12}=V(\mathbf{r}_{(1)}-\mathbf{r}_{(2)}),$ the
prefactor $V_{\{\mathbf{p}_{(n)}\}\mathbf{k}}$ reads
\begin{equation}
V_{\{\mathbf{p}_{(n)}\}\mathbf{k}}^{(b)}=\frac{2C_{\psi }}{(2\pi )^{9/2}}V(%
\mathbf{p}_{(1)}-\mathbf{k})\cos [\mathbf{\mathcal{P}}(t)\cdot \mathbf{R}%
/2]\varphi _{0}^{(2)}(\mathbf{\mathcal{P}}(t))  \label{Vpnkbond}
\end{equation}%
or%
\begin{equation}
V_{\{\mathbf{p}_{(n)}\mathbf{\}k}}^{(a)}=\frac{2iC_{\psi }}{(2\pi )^{9/2}}V(%
\mathbf{p}_{(1)}-\mathbf{k})\sin [\mathbf{\mathcal{P}}(t)\cdot \mathbf{R}%
/2]\varphi _{0}^{(2)}(\mathbf{\mathcal{P}}(t)),  \label{Vpnkanti}
\end{equation}%
with $\mathbf{\mathcal{P}}(t)=\mathbf{\tilde{p}}_{(1)}(t)+\mathbf{\tilde{p}}%
_{(2)}(t)-\mathbf{\tilde{k}}(t)$, for bonding and antibonding orbitals,
respectively. Thereby,
\begin{equation}
\varphi _{0}^{(2)}(\mathbf{\mathcal{P}}(t))=\int d^{3}r_{(2)}\exp [i\mathbf{%
\mathbf{\mathcal{P}}}(t)\cdot \mathbf{r}_{(2)})]\phi _{0}^{(2)}(\mathbf{r}%
_{(2)}),
\end{equation}%
and

\begin{equation}
V(\mathbf{p}_{(1)}-\mathbf{k})=\int d^{3}rV(\mathbf{r})\exp [i(\mathbf{p}%
_{(1)}-\mathbf{k})\cdot \mathbf{r}],
\end{equation}%
with $\mathbf{r=r}_{(1)}-\mathbf{r}_{(2)}.$ In the velocity and length
gauges, the argument in Eqs. (\ref{Vpnkbond}), (\ref{Vpnkanti}) is given by $%
\mathbf{\mathcal{P}}(t)=\mathbf{p}_{(1)}+\mathbf{p}_{(2)}-\mathbf{k}$ and $%
\mathbf{\mathcal{P}}(t)=\mathbf{p}_{(1)}+\mathbf{p}_{(2)}-\mathbf{k+A}(t)$,
respectively.

In terms of the momentum components $p_{(n)\parallel },$ or $p_{(n)\perp }$ $%
(n=1,2),$ parallel and perpendicular to the laser-field polarization,
condition (\ref{Vpnkbond}) or (\ref{Vpnkanti}) may be written as $\cos
[(\zeta _{\parallel }+\zeta _{\perp })R/2]$ or $\sin [(\zeta _{\parallel
}+\zeta _{\perp })R/2]$, respectively, with%
\begin{equation}
\zeta _{\parallel }=\left[ \sum\limits_{i=1}^{2}p_{(n)\parallel }-\kappa (t)%
\right] \cos \theta ,  \label{argpar}
\end{equation}%
and%
\begin{equation}
\mathbf{\zeta }_{\perp }=p_{(1)\perp }\sin \theta \cos \varphi +p_{(2)\perp
}\sin \theta \cos (\varphi +\alpha ).  \label{argperp}
\end{equation}%
Thereby, the term $\kappa (t)$ is equal to $k-A(t)$ in the length gauge and
to $k$ in the velocity gauge. Eq.~(\ref{argpar}) provides well-defined
interference fringes, as functions of the parallel momenta ($p_{(1)\parallel
},p_{(2)\parallel })$. Eq.~(\ref{argperp}), on the other hand, has no
obvious dependence on the alignment angle $\theta $. Indeed, because it
depends on the angles $\varphi $ and $\alpha ,$ in the momentum plane
spanned by the perpendicular momentum components, when one integrates over
such variables, it main effect is to is blur the interference fringes. In
this work, we will consider parallel-aligned molecules. This implies that
Eq.~(\ref{argperp}) vanishes, and therefore that the interference patterns
are sharpest. Explicitly, the interference conditions will be given by
\begin{equation}
p_{(1)\parallel }+p_{(2)\parallel }=\frac{N\pi }{R}+\kappa (t),
\label{fringespar}
\end{equation}%
where $N$ is an integer. For a symmetric combination of atomic orbitals,
even or odd $N$ gives the interference maxima and the minima, respectively,
while in the antisymmetric case the situation is reversed.

\subsection{Modified saddle-point equations}

\label{saddmodified}

We will now incorporate the structure of the molecule, which is embedded in
the prefactors (\ref{Vk0bond})-(\ref{Vpnkanti}), in the semiclassical
action. Subsequently, the transition amplitudes obtained will be computed
employing a uniform saddle-point approximation (c.f. \cite{atiuni} for
details). For that purpose, we will exponentialize the prefactors $V_{%
\mathbf{k}0}$ and $V_{\{\mathbf{p}_{(n)}\}\mathbf{k}}$. This procedure
allows one to single out different rescattering scenarios, involving one or
two centers.

This yields the sum%
\begin{equation}
M=\sum_{j=1}^{2}\sum_{\nu =1}^{2}M_{j\nu }  \label{sumampl}
\end{equation}%
of the transition amplitudes%
\begin{equation}
M_{j\nu }=\int\limits_{-\infty }^{\infty }dt\int\limits_{-\infty
}^{t}dt^{\prime }\int d^{3}k\varphi _{0}^{(2)}(\mathbf{\mathcal{P}}(t))%
\mathcal{I}(\mathbf{\tilde{k}}(t^{\prime }))\exp [iS_{j\nu }(\{\mathbf{p}%
_{(n)}\},\mathbf{k},t,t^{\prime })],
\end{equation}%
with the modified action $S_{j\nu }(\{\mathbf{p}_{(n)}\},\mathbf{k}%
,t,t^{\prime })$ defined as%
\begin{equation}
\hspace*{-0.5cm}\hspace*{-0.5cm}S_{j\nu }(\{\mathbf{p}_{(n)}\},\mathbf{k}%
,t,t^{\prime })=S(\{\mathbf{p}_{(n)}\},\mathbf{k},t,t^{\prime })+(-1)^{\nu }%
\hspace{-0.1cm}\left[ \mathbf{\mathcal{P}}(t)\hspace{-0.1cm}+\hspace{-0.1cm}%
(-1)^{j+\nu }\mathbf{\tilde{k}}(t^{\prime })\right] \cdot \frac{\mathbf{R}}{2%
},  \label{ssame}
\end{equation}%
where the indices $j,\nu $ relate to the centers in the molecule. In the
above-stated equation, for processes involving a single center, $j+\nu $ is
even, while for scattering scenarios in which the first electron leaves at a
center $C_{j}$ and rescatters with a center $C_{\nu },\nu \neq j$, $j+\nu $
is odd$.$ The saddle-point equations $\partial _{t}S_{j\nu }(\{\mathbf{p}%
_{(n)}\},\mathbf{k},t,t^{\prime })=\partial _{t^{\prime }}S_{j\nu }(\{%
\mathbf{p}_{(n)}\},\mathbf{k},t,t^{\prime })=0$ and $\partial _{\mathbf{k}%
}S_{j\nu }(\{\mathbf{p}_{(n)}\},\mathbf{k},t,t^{\prime })=\mathbf{0}$ are
explicitly given by%
\begin{equation}
\frac{\left[ \mathbf{k}+\mathbf{A}(t^{\prime })\right] ^{2}}{2}%
=-E_{0(1)}+(-1)^{2\nu +j+1}\partial _{t^{\prime }}\mathbf{\tilde{k}}%
(t^{\prime })\cdot \mathbf{R}/2,  \label{saddmod1}
\end{equation}

\begin{equation}
\int_{t^{\prime }}^{t}d\tau \left[ \mathbf{k}+\mathbf{A}(\tau )\right]
+(-1)^{\nu }\partial _{\mathbf{k}}\left\{ \left[ (-1)^{j+\nu }\mathbf{\tilde{%
k}}(t^{\prime })-\mathbf{\tilde{k}}(t)\right] \cdot \mathbf{R}/2\right\} =0,
\label{saddmod2}
\end{equation}%
and
\begin{equation}
\sum_{n=1}^{2}\frac{[\mathbf{p}_{(n)}+\mathbf{A}(t)]^{2}}{2}=\frac{\left[
\mathbf{k}+\mathbf{A}(t)\right] ^{2}}{2}-E_{0(2)}+(-1)^{\nu +1}\partial _{t}%
\mathbf{\mathcal{P}}(t)\hspace{-0.1cm}\cdot \mathbf{R}/2.  \label{saddmod3}
\end{equation}%
Eq. (\ref{saddmod1}) expresses the conservation of energy at $t^{\prime },$
with tunneling ionization of the first electron. Eq.~(\ref{saddmod2})
provides the condition for the first electron to return, either to the site
of its release or to the other ion. Finally, Eq.~(\ref{saddmod3}) yields the
conservation of energy at the instant of rescattering.

One should note that the saddle-point equations (\ref{saddmod1}) and (\ref%
{saddmod3}) are gauge dependent. Specifically, in the length gauge, the
tunneling and rescattering conditions are given by%
\begin{equation}
\frac{\left[ \mathbf{k}+\mathbf{A}(t^{\prime })\right] ^{2}}{2}%
=-E_{0(1)}+(-1)^{2\nu +j}\mathbf{E}(t^{\prime })\cdot \mathbf{R}/2
\label{saddlen1}
\end{equation}%
and%
\begin{equation}
\sum_{n=1}^{2}\frac{[\mathbf{p}_{(n)}+\mathbf{A}(t)]^{2}}{2}=\frac{\left[
\mathbf{k}+\mathbf{A}(t)\right] ^{2}}{2}-E_{0(2)}+(-1)^{\nu }\mathbf{E}%
(t)\cdot \mathbf{R}/2  \label{saddlen3}
\end{equation}%
respectively. Physically, the additional terms in Eqs.~(\ref{saddlen1}) and (%
\ref{saddlen3}) may be interpreted as potential-energy shifts, which sink or
increase the ionization potentials. For instance, for single-center
processes $(j=\nu ),$ such shifts are symmetric, whereas for two-center
scattering scenarios $(j\neq \nu )$ they possess the same sign. In the
velocity gauge, the corresponding equations read
\begin{equation}
\left[ \mathbf{k}+\mathbf{A}(t^{\prime })\right] ^{2}=-2E_{0(1)},
\label{saddle1}
\end{equation}%
and
\begin{equation}
\sum_{n=1}^{2}\frac{[\mathbf{p}_{(n)}+\mathbf{A}(t)]^{2}}{2}=\frac{\left[
\mathbf{k}+\mathbf{A}(t)\right] ^{2}}{2}-E_{0(2)},  \label{saddle3}
\end{equation}%
respectively. Eq.~(\ref{saddle1}) and (\ref{saddle2}) are identical to the
NSDI saddle-point equations for a single atom expressing tunneling
ionization at $t^{\prime }$ and conservation of energy at $t$. Furthermore,
in the limit $E_{0(1)}\rightarrow 0$, Eq. (\ref{saddle1}) may be interpreted
as \ the initial kinetic energy of a classical particle, which starts its
motion with vanishing drift momentum.

If written in terms of the electron momentum components parallel and
perpendicular to the laser-field polarization, the rescattering conditions (%
\ref{saddlen3}) and (\ref{saddle3}) read%
\begin{equation}
\sum_{n=1}^{2}\frac{[p_{(n)||}+A(t)]^{2}}{2}+\frac{\mathbf{p}_{(n)\perp }^{2}%
}{2}=\frac{\left[ \mathbf{k}+\mathbf{A}(t)\right] ^{2}}{2}-\tilde{E}_{0(2)}.
\label{hyper}
\end{equation}%
In the length and velocity gauges, $\tilde{E}_{0(2)}=E_{0(2)}+(-1)^{\nu }%
\mathbf{E}(t)\cdot \mathbf{R}/2$ and $\tilde{E}_{0(2)}=E_{0(2)}$,
respectively. In the former case, $\tilde{E}_{0(2)}$ can be viewed as an
effective second ionization potential. The above-stated expression gives the
equation of a six-dimensional hypersphere in momentum space, whose radius
corresponds to the momentum region for which electron-impact ionization has
a classical counterpart. If the kinetic energy of the first electron, upon
return, is smaller than $\tilde{E}_{0(2)}$, then the second electron cannot
be released and this process is classically forbidden. In this case, the
corresponding transition probability is vanishingly small.

The return condition, and how it is related to such indices, can be clearly
seen in Eq.~(\ref{saddmod2}). In fact, if $j=\nu $, the additional terms
vanish, and the condition
\begin{equation}
\int_{t^{\prime }}^{t}d\tau \left[ \mathbf{k}+\mathbf{A}(\tau )\right] =0,
\label{saddle2}
\end{equation}%
is obtained. Eq.~(\ref{saddle2}) constrains the value of the intermediate
momentum $\mathbf{k}$ so that the electron is leaving and returning to the
same center in the molecule. On the other hand, if $j\neq \nu ,$ Eq.~(\ref%
{saddmod2}) reads%
\begin{equation}
\int_{t^{\prime }}^{t}d\tau \left[ \mathbf{k}+\mathbf{A}(\tau )\right]
+(-1)^{\nu +1}\mathbf{R}=0.  \label{sadtwocv}
\end{equation}%
The negative and the positive sign corresponds to the situation in which the
electron starts from the center $C_{1}$ and rescatters with $C_{2},$ or
starts from $C_{2}$ and rescatters with $C_{1},$ respectively. Both Eq. (\ref%
{saddle2}) and Eq. (\ref{saddmod2}) are identical to the return conditions
obtained from the classical equations of motion of an electron starting its
orbit with vanishing drift momentum, and propagating under the influence of
solely the laser field.

One should note that, within the context of high-order harmonic generation
and above-threshold ionization, the additional potential-energy shifts
present in the length-gauge SFA have raised a great deal of controversy.
Indeed, there has been considerable debate whether such shifts are not an
artifact of the strong-field approximation. Their existence is mainly
related to the fact that the ions in the molecule are at a distance $\pm
\mathbf{R}/2$ from the origin of the coordinate system, which is located at
the geometric center of the molecule \cite{dressedSFA,BCM2007}. Furthermore,
their presence does not allow an immediate connection with the classical
equations of motion of an electron in an external laser field, as in the
velocity-gauge formulation. For the above-stated reasons, it was suggested
that these additional phase shifts should be removed by dressing the
electronic bound states \cite{BCM2007}.

In recent HHG computations, however, it has been found that, if the
additional potential-energy shifts are removed by employing field-dressed
states, the two-center interference patterns break down \cite%
{F2007,dressedSFA2}. A direct comparison with the numerical solution of the
time-dependent Schr\"{o}dinger equation has shown that this should only
occur for very large internuclear distances, for which the atoms
constituting the molecule practically behave as isolated entities \cite%
{LC2008}. Unfortunately, for NSDI, there are neither realistic computations
by other means nor enough experiments to settle the issue. Therefore, in the
following, we will compute electron momentum distributions in both length
and velocity gauges without favoring any of them.

\section{Electron momentum distributions}

\label{results}

We will approximate the external laser field by a monochromatic wave $%
\mathbf{E}(t)=\varepsilon _{0}\sin \omega t\mathbf{e}_{x}.$ In this case,
the electron momentum distributions, as functions of the momentum components
$(p_{(1)\parallel },p_{(2)\parallel })$ parallel to the laser-field
polarization, read%
\begin{equation}
F(p_{(1)\parallel },p_{(2)\parallel })=\int \hspace*{-0.2cm}\int
d^{2}p_{(1)\perp }d^{2}p_{(2)\perp }|M_{R}(\mathbf{p}_{(1)},\mathbf{p}%
_{(2)})+M_{L}(\mathbf{p}_{(1)},\mathbf{p}_{(2)})|^{2},  \label{distr}
\end{equation}%
where the indices\ $R$ and $L$ denote the right and left peaks,
respectively. The amplitude $M_{R}$ is given by Eq. (\ref{Mp}), and $M_{L}(%
\mathbf{p}_{(1)},\mathbf{p}_{(2)})=M_{R}(-\mathbf{p}_{(1)},-\mathbf{p}%
_{(2)}) $. In $M_{L}(\mathbf{p}_{(1)},\mathbf{p}_{(2)})$, the action, the
field and prefactors are displaced by half a cycle with respect to $M_{R}$,
i.e., we used the symmetry $\mathbf{A}(t)=\pm \mathbf{A}(t\pm T/2)$.

The above-mentioned symmetry will guarantee that the overall electron
momentum distributions, obtained either using the prefactor (\ref{Vk0bond})-(%
\ref{Vpnkanti}) or all four different transition amplitudes, will be
symmetric with regard to the simultaneous inversion $(p_{(1)||},p_{(2)||})%
\rightarrow (-p_{(1)||},-p_{(2)||})$ of the parallel momentum components.
Specifically for the transition amplitudes $M_{j\nu }$, we observe that $%
M_{j\nu }(\mathbf{p}_{(1)},\mathbf{p}_{(2)})=M_{\nu j}(-\mathbf{p}_{(1)},-%
\mathbf{p}_{(2)})$. If both start and rescattering times $t,t^{\prime }$
were real, this would lead to symmetric distributions upon $%
(p_{(1)||},p_{(2)||})\rightarrow (-p_{(1)||},-p_{(2)||}).$ Furthermore, this
would also guarantee that $|M_{11}+M_{12}|^{2}=|M_{22}+M_{21}|^{2}$ and $%
|M_{12}+M_{22}|^{2}=|M_{11}+M_{21}|^{2}$. Since, however, the first electron
is released through tunneling ionization, $\mathrm{Im}[t^{\prime }]\neq 0$
and this is not necessarily so. This issue is particularly important in the
length-gauge formulation of the SFA, and will be discussed subsequently.

In the form factor $V_{\mathbf{k}0}$, we considered Coulomb-type binding
potentials $V_{0}$ and a symmetric combination of $1s$ orbitals. We assume
that the second electron is dislodged by a contact-type interaction $V_{12}$
placed at the position of the ions. Explicitly,
\begin{equation}
V_{12}=\delta (\mathbf{r}_{(1)}-\mathbf{r}_{(2)})\left[ \delta (\mathbf{r}%
_{(2)}-\mathbf{R}/2)+\delta (\mathbf{r}_{(2)}+\mathbf{R}/2)\right] ,
\label{contact}
\end{equation}%
and can be viewed as an effective interaction, which in a rough way,
accounts for the presence of the residual ions. This interaction has been
employed in Ref. \cite{FSLY2008}, within a two-center context, and has also
been widely used in the single-atom case \cite{FSLB2004,FL2005}. In the
specific context of this work, it has the advantage of eliminating any
additional momentum dependence from $\varphi _{0}^{(2)}(\mathcal{P}(t))$,
whose effect could be hard to disentangle from the two-center interference.

\begin{figure}[tbp]
\begin{center}
\includegraphics[width=9cm]{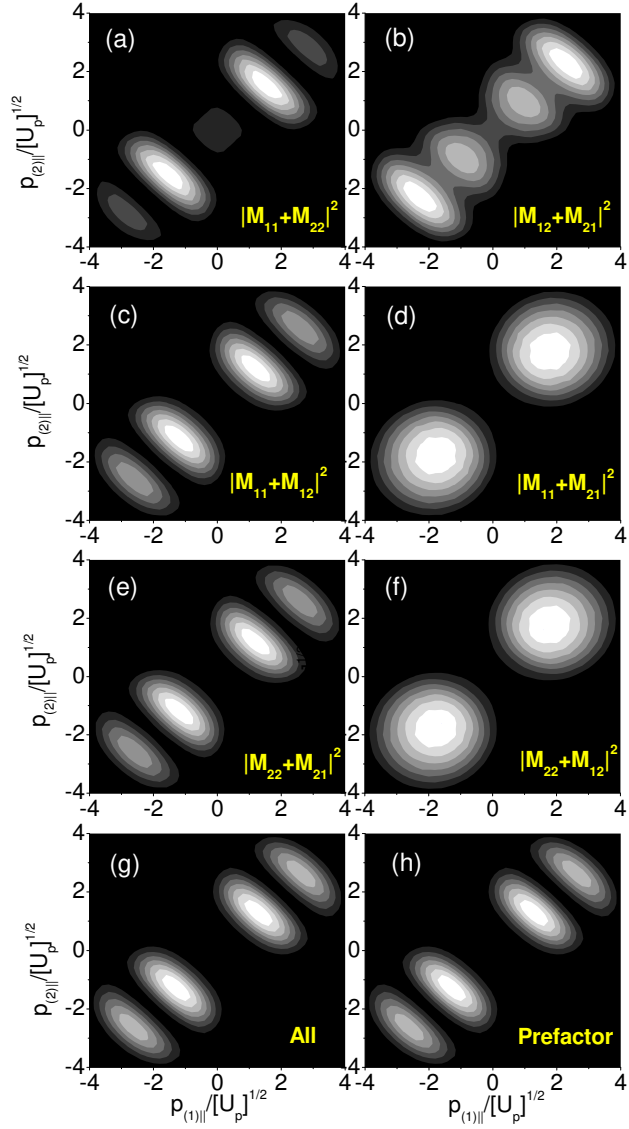}
\end{center}
\caption{Contributions from different scattering scenarios to the electron
momentum distributions as functions of the momentum components $%
(p_{(1)\parallel },p_{(2)\parallel })$ parallel to the laser-field
polarization. The distributions have been computed in the velocity gauge and
for a symmetric combination of 1s orbitals. The field intensity and
frequency have been taken as $I=1.5\times 10^{14}\mathrm{W/cm}^{2}$, and $%
\protect\omega =0.057$ a.u., respectively, and the ionization potentials $%
E_{01}=0.57$ a.u. and $E_{02}=0.98$ a.u. correspond to $N_{2}$ at the
equilibrium internuclear distance $R=2.068$ a.u. The upper panels display
the contributions from topologically similar scattering scenarios, involving
only one or two centers, i.e., the transition probabilities $%
|M_{11}+M_{22}|^{2}$ and $|M_{12}+M_{21}|^{2}$[panels (a) and (b),
respectively]. In panels (c) and (d), we display the contributions from the
processes starting and ending at center $C_{1}$, respectively (transition
probabilities $|M_{11}+M_{12}|^{2}$ and $|M_{11}+M_{21}|^{2},$
respectively). Panels (e) and (f) depict the contributions $%
|M_{21}+M_{22}|^{2}$ and $|M_{12}+M_{22}|^{2}$ from those starting and
ending at $C_{2},$ respectively. The sum $|M_{11}+M_{12}+M_{21}+M_{22}|^{2}$
of all contributions are provided in panel (g). For comparison, panel (h)
has been computed using the symmetric prefactors (\protect\ref{Vk0bond}) and
(\protect\ref{Vpnkbond}) and single-atom saddle-point equations.}
\label{processesv}
\end{figure}

Throughout, we employ a higher driving-field intensity than those reported
in typical NSDI experiments involving diatomic molecules \cite%
{NSDIsymm,NSDIalign}. This has been done with the purpose of extending the
region in momentum space for which electron-impact ionization exhibits a
classical counterpart, i.e., the radius of the hypersphere (\ref{hyper}). In
previous work, we have shown that this was necessary in order to make a
detailed assessment of quantum-interference effects in this context \cite%
{FSLY2008}. Furthermore, we restrict ourselves to the case of a symmetric
combination of $1s$ orbitals. This gives, for the specific interaction (\ref%
{contact}),
\begin{equation}
V_{\{\mathbf{p}_{(n)}\}\mathbf{k}}\sim \cos [\mathbf{\mathcal{P}}(t)\cdot
\mathbf{R}/2]\psi _{100}(0)
\end{equation}%
Without loss of generality, however, our studies, as well as the conclusions
of this work, may be extended to the antisymmetric case, or to more complex
orbitals. In particular, if the interaction (\ref{contact}) is taken, only\
the contributions from s states would be non-vanishing in the form factor $%
V_{\{\mathbf{p}_{(n)}\}\mathbf{k}}.$

In Fig.~\ref{processesv}, we display the electron momentum distributions in
the velocity gauge and for symmetric orbitals, considering specific
scattering processes. For comparison, in Fig.~\ref{processesv}.(h), we are
presenting the distributions obtained employing the prefactors (\ref{Vk0bond}%
) and (\ref{Vpnkbond}) and solving single-atom saddle point equations. This
approach has been \ considered in \cite{FSLY2008}, and incorporates all the
structure of the molecule in the prefactors. This distribution exhibits
fringes in agreement with the interference condition (\ref{fringespar}).

In Fig.~\ref{processesv}.(a), only the contributions from the processes in
which the first electron leaves and returns to the same center have been
taken into account. These processes are related to the transition amplitudes
$M_{jj}(j=1,2).$ In this case, fringes parallel to the anti-diagonal $%
p_{(1)||}=-p_{(2)||}$ are also present. Their position, however, disagrees
with condition (\ref{fringespar}). A closer inspection, however, shows that
the terms $M_{11}$ and $M_{22}$ can be grouped as $\cos [(\mathbf{p}_{(1)}+%
\mathbf{p}_{(2)})\cdot \mathbf{R}/2].$ This gives another interference
condition in terms of the parallel momentum components, namely $%
p_{(1)||}+p_{(2)||}=N\pi /R$. Maxima and minima are present for even and odd
$N,$ respectively. We have verified that all maxima and minima in the figure
are approximately given by the above-stated expression, and correspond to
the integers $0\leq N\leq 4$.

If only the two-center processes are taken [Fig.~\ref{processesv}.(b)], the
transition amplitudes $\ M_{21}$ and $M_{12}$ can be grouped as $\cos [(%
\mathbf{p}_{(1)}+\mathbf{p}_{(2)}-2\mathbf{k})\cdot \mathbf{R}/2]$. This
gives fringes following $p_{(1)||}+p_{(2)||}-2k=N\pi /R,$ which, once more,
differ from the global interference condition (\ref{fringespar}). A rough
analytic estimate of the position of the fringes can be made, by considering
that the first electron leaves at peak field and returns at a field
crossing. In this estimate, instead of using the modified return conditions (%
\ref{sadtwocv}), we considered the saddle-point equation (\ref{saddle2}),
which states that the electron returns to the site of its release. In the
present context, this expression for $k$ may also be viewed as an average
value between those given by the two-center return conditions given in (\ref%
{sadtwocv}). For the specific parameters in this work, this yields $%
p_{(1)||}+p_{(2)||}\simeq (1.45N-0.849)\sqrt{U_{p}}.$ This estimate is in
good agreement with the maxima and the minima displayed in Fig.~\ref%
{processesv}.(b).

Interestingly, the distributions obtained from the transition probability $%
|M_{jj}+$ $M_{j\nu }|^{2},$ with $j=1,2$ and $j\neq \nu ,$ agree with the
overall interference condition (\ref{fringespar}). This is shown in Figs.~%
\ref{processesv}.(c) and \ref{processesv}.(e), and is due to the fact that
one may rewrite the sum of such terms as $\exp [\pm i\mathbf{k}\cdot \mathbf{%
R}/2]\cos [(\mathbf{p}_{(1)}+\mathbf{p}_{(2)}-\mathbf{k})\cdot \mathbf{R}%
/2]. $ Apart from an overall phase factor, this is the same interference
condition as if all transition amplitudes are taken. This latter case is
shown in Fig.~\ref{processesv}.(g), for comparison. Physically, the term $%
M_{jj}$ corresponds to the orbits in which the first electron is freed at
and subsequently rescatters off a specific center $C_{j},$ while $M_{j\nu }$
gives the process in which it reaches the continuum at $C_{j}$ and returns
to the other center $C_{\nu }.$ This suggests that these processes are the
most relevant in determining the interference patterns from NSDI in a
molecule.
\begin{figure}[tbp]
\begin{center}
\includegraphics[width=9cm]{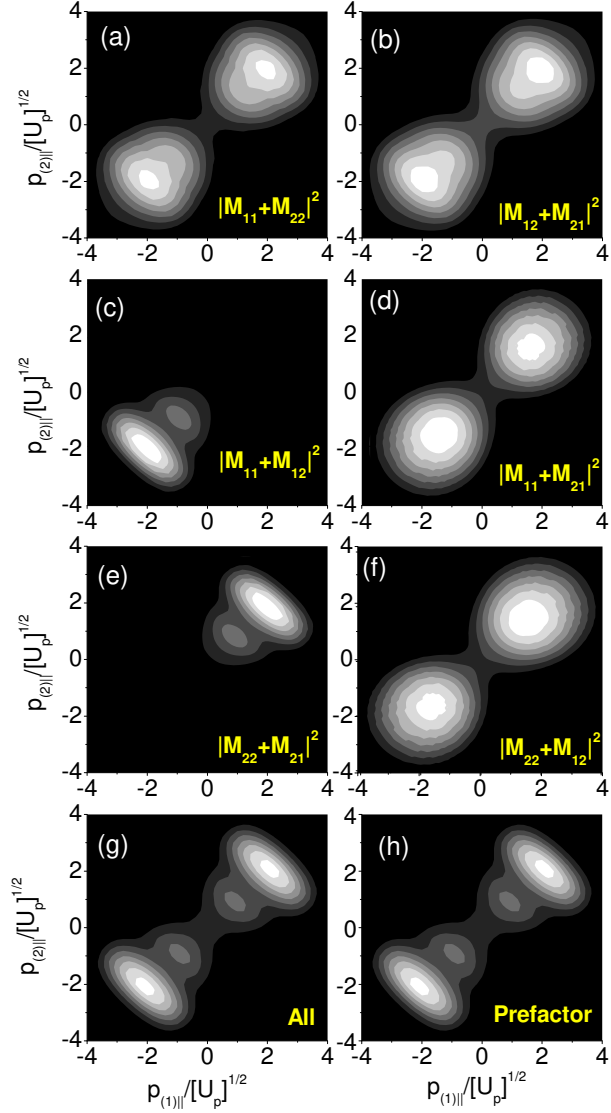}
\end{center}
\caption{Contributions from different scattering scenarios to the electron
momentum distributions as functions of the momentum components $%
(p_{(1)\parallel },p_{(2)\parallel })$ parallel to the laser-field
polarization. The distributions have been computed in the length gauge and
for a symmetric combination of 1s orbitals. The remaining parameters are the
same as in the previous figure. The upper panels display the contributions
from topologically similar scattering scenarios, involving only one or two
centers, i.e., the transition probabilities $|M_{11}+M_{22}|^{2}$ and $%
|M_{12}+M_{21}|^{2}$ [panels (a) and (b), respectively]. In panels (c) and
(d), we display the contributions from the processes starting and ending at
center $C_{1}$, respectively (transition probabilities $|M_{11}+M_{12}|^{2}$
and $|M_{11}+M_{21}|^{2},$ respectively). Panels (e) and (f) depict the
contributions $|M_{21}+M_{22}|^{2}$ and $|M_{12}+M_{22}|^{2}$ from those
starting and ending at $C_{2},$ respectively. The sum $%
|M_{11}+M_{12}+M_{21}+M_{22}|^{2}$ of all contributions are provided in
panel (g). For comparison, panel (h) has been computed using the prefactors (%
\protect\ref{Vk0bond}) and (\protect\ref{Vpnkbond}) and the single-atom
saddle-point equations.}
\label{processesl}
\end{figure}

On the other hand, if we consider scattering scenarios ending at the same
center, the electron-momentum distributions resemble very much those
obtained for the single-atom case, namely isotropic distributions centered
near $p_{(1)||}=p_{(2)||}=\pm 2\sqrt{U_{p}}$. This holds regardless of
whether one or two centers are involved, and can be seen in Figs.~\ref%
{processesv}.(d) and \ref{processesv}.(f). In each panel, both one and two
center processes are considered. However, interference fringes are absent.
This is consistent with the double-slit physical picture, which relates the
existence of the interference maxima and minima to photoelectron emission in
spatially separated centers. This can also be seen by rewriting these
contributions as $\exp [\pm i(\mathbf{p}_{(1)}+\mathbf{p}_{(2)}-\mathbf{k}%
\cdot \mathbf{R}/2)]V_{\mathbf{k}0}.$ The form factor $V_{\mathbf{k}0}$ does
not lead to well-defined interference patterns, so that the distributions
are very similar to those obtained in the single-atom case (for discussions
see \cite{FSLY2008,NSDIsymm}). Finally, we observe that Figs.~\ref%
{processesv}.(d) and \ref{processesv}.(f) are practically identical. This
means that the condition $|M_{11}+M_{21}|^{2}=$ $|M_{22}+M_{12}|^{2}$ holds%
\footnote{%
We have verified, by a series of numerical computations, that this
condition only holds in practice. This is a consequence of the fact
that the start time $t^{\prime }$ exhibits a non-vanishing imaginary
time, and that, according to the saddle-point Eqs. (29), (30), the
intermediate momentum depends on this variable. The relative
differences in the yields, however, is less than $0.2\%.$}.
Intuitively, this is expected, as the transition amplitudes
$M_{j1}(j=1,2)$ are the complex conjugates of $M_{j2}.$ The same
holds for Figs.~\ref{processesv}.(c) and \ref{processesv}.(e), which
expresses the fact that $|M_{11}+M_{12}|^{2}=$
$|M_{22}+M_{21}|^{2}$.

The length-gauge counterparts of the distributions discussed above are
presented in Fig.~\ref{processesl}. In this case, we expect a different
interference condition, according to Eq.~(\ref{fringespar}). Furthermore, an
inspection of the saddle-point equations shows that there exist additional
potential-energy terms which lower or increase the barrier through which the
first electron tunnels out. \ These energy shifts will influence the
electron-momentum distributions, as we will discuss subsequently.

In the upper panels, we consider topologically similar scattering scenarios,
involving only one or two centers [Figs.~\ref{processesl}.(a) and \ref%
{processesl}.(b), respectively]. In contrast to the velocity-gauge
situation, these distributions exhibit at most a slight distortion, as
compared to the single-atom case. Sharp interference fringes, however, are
absent. This is a consequence of the above-mentioned potential-energy
shifts. Such shifts alter the potential-energy barriers in such a way that
is much more probable for the first electron to tunnel out from a specific
center of the molecule. Hence, the processes starting from a center $C_{j}$
in the molecule will be far more prominent than those starting from the
other center $C_{\nu }(\nu \neq j)$ and there will be no sharp fringes. For
the parameters employed in this figure, we estimate a difference of roughly
one order of magnitude between each transition amplitude. A similar effect
was present for high-order harmonic generation and has been discussed in
detail in Ref.~\cite{F2007}.

Also in the length gauge, the transition amplitudes $M_{jj}+M_{j\nu }$
related the processes starting at the same center and ending at different
centers lead to the same interference condition as the overall prefactors (%
\ref{Vk0bond}) and (\ref{Vpnkbond}), apart from a phase factor $\exp [\pm
i\xi ],$ with $\xi =[\mathbf{k}+\mathbf{A}(t^{\prime })]\cdot \mathbf{R}/2.$
There are, however, some major quantitative differences, due to the fact
that $\mathrm{Im}[t^{\prime }]\neq 0$. This leads to a non-vanishing
imaginary part in $\xi $, which, depending on the field, will enhance the
yield for a center $C_{j}$ and suppress it for the other center $C_{\nu }$
in the molecule. As the field reverses its sign, the contributions from $%
C_{j}$ will be suppressed and those from the other center enhanced. This
leads to strongly asymmetric electron momentum distributions, and may be
interpreted as another consequence of the additional potential-energy
shifts, which are present only in the length-gauge formulation of the SFA.

The above-stated features are explicitly shown in Figs.~\ref{processesl}.(c)
and \ref{processesl}.(e), starting from $C_{1}$ and $C_{2},$ respectively.
In the former case [Figs.~\ref{processesl}.(c)], the yield is strongly
suppressed in the region of positive parallel momentum. This is due to the
fact that the additional potential energy shifts suppresses tunneling
ionization for all the orbits starting from $C_{1}$. As the field reverses
its sign, i.e., when $t^{\prime }\rightarrow t^{\prime }+T/2$, the barrier
will be sunk by these shifts. This explains the much larger yield in the
negative momentum region. Fig. \ref{processesl}.(e), which is the mirror
image of Figs.~\ref{processesl}.(c) with respect to $(p_{1||},p_{2||})%
\rightarrow (-p_{1||},-p_{2||}),$ exhibits the consequences of the opposite
effect: for the time $t^{\prime }$, the barrier is sunk for $C_{2}.$ This
leads to enhanced distributions in the positive momentum regions. After half
a cycle, as the field changes its signs, the barrier will be lifted and and
the contributions in the negative momentum region suppressed. A similar
effect has been observed for high-order harmonic generation \cite%
{F2007,F2008}, In particular, in \cite{F2008}, apart from modified
saddle-point equations, we have employed effective prefactors and
single-atom saddle-point equations in order to mimic the interference
between different processes. Also in the latter case, we observed different
orders of magnitude for processes starting at different centers.

Therefore, in contrast to the velocity-gauge situation, the processes
starting at a particular center of the molecule do not lead to the same
distributions as if all four terms in Eq. (\ref{ssame}) are taken. These
latter results are displayed in Fig.~\ref{processesl}.(g). For comparison,
in Fig.~\ref{processesl}.(h), we also provide the outcome of the
computations using single-atom saddle-point equations and the prefactor (\ref%
{Vpnkbond}) and (\ref{Vk0bond}). Finally, if the processes starting at
different centers, but ending at the same center lead to distributions
qualitatively similar than those observed in the single-atom case [see Fig. %
\ref{processesl}.(d) and (f)]. Similarly to the velocity-gauge situation,
the terms can be combined in $\exp [\pm i(\mathbf{p}_{(1)}+\mathbf{p}_{(2)}-(%
\mathbf{k-A}(t)\mathbf{)}\cdot \mathbf{R}/2)]\cos [\mathbf{k+A}(t^{\prime
})].$ Even if the extra dependence on the vector potential influences the
distributions, qualitatively, similar conclusions hold. However, there is a
slight asymmetry with respect to the reflection $(p_{(1)||},p_{2||})%
\rightarrow (-p_{(1)||},-p_{(2)||})$ due to the fact that the times $%
t,t^{\prime }$ are complex. Once more, the contributions of the processes
ending at a particular center $C_{j}$ are the mirror image those ending at
the other center $C_{\nu }$.
\begin{figure}[tbp]
\begin{center}
\includegraphics[width=9cm]{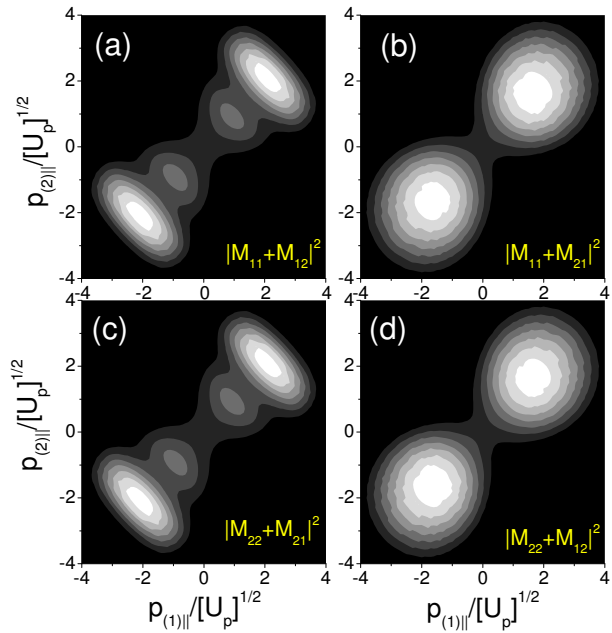}
\end{center}
\caption{Transition probabilities starting and ending at the same centers,
considering only the real parts of $t,t^{\prime },$ as functions of the
momentum components $(p_{(1)\parallel },p_{(2)\parallel })$ parallel to the
laser-field polarization, for the same parameters as in Fig. 2. Panels (a),
and (b) exhibit the contributions from the processes starting and ending at
center $C_{1}$ (transition probabilities $|M_{11}+M_{12}|^{2}$ and $%
|M_{11}+M_{21}|^{2},$ respectively), while panels (c) and (d) depict the
contributions $|M_{21}+M_{22}|^{2}$ and $|M_{12}+M_{22}|^{2}$ from those
starting and ending at $C_{2},$ respectively. }
\end{figure}

In the following, we explicitly show that the above-mentioned asymmetries
are caused by $\mathrm{Im}[t]$, and $\mathrm{Im}[t^{\prime }].$ For that
purpose, in Fig. 3, we plot the counterparts of Fig. 2(c)-2(f) employing
only the real parts of such times. In this case, the contributions of the
orbits starting at the same center are symmetric, i.e., the condition $%
|M_{11}+M_{12}|^{2}=|M_{22}+M_{21}|^{2}$ holds [see Figs. 3.(a) and 3.(c)].
The contributions ending at the same center, depicted in Figs. 3.(b) and
3.(d), also exhibit the symmetry $|M_{11}+M_{21}|^{2}=|M_{22}+M_{12}|^{2}.$
In these computations, instead of employing modified saddle-point equations,
we grouped the separate transition amplitudes in the effective prefactors
%%%%%%%%%%%%%%%%%%%%%%
\begin{eqnarray}
V_{C_{j}}^{\mathrm{start}}(\{\mathbf{p}_{(n)}\},\mathbf{k}) &\sim &\exp
[i(-1)^{j}[\mathbf{k}+\mathbf{A}(t^{\prime })]\cdot \mathbf{R}/2] \\
&&\times \cos [(\mathbf{p}_{(1)}+\mathbf{p}_{(2)}-\mathbf{k+A}(t))\cdot
\mathbf{R}/2]  \nonumber
\end{eqnarray}%
and %%%%%%%%%%%%%%%%%%%%%%%%%%%
\begin{eqnarray}
V_{C_{j}}^{\mathrm{end}}(\{\mathbf{p}_{(n)}\},\mathbf{k}) &\sim &\exp
[i(-1)^{j}(\mathbf{p}_{(1)}+\mathbf{p}_{(2)}-\mathbf{k+A}(t)\cdot \mathbf{R}%
/2)] \\
&&\times \cos [\mathbf{k+A}(t^{\prime })]  \nonumber
\end{eqnarray}%
mimicking the processes starting and ending at a center $C_{j}$,
respectively, and employed single-atom saddle-point equations (see Ref. \cite%
{F2008} for details on this procedure). The imaginary parts of $t,t^{\prime }
$ were removed in the prefactors only.

\section{Conclusions}

\label{concl}

We have made a detailed analysis of the different scattering scenarios in
laser-induced nonsequential double ionization of diatomic molecules, within
the framework of the strong-field approximation. We assumed that the second
electron is dislodged by electron-impact ionization and employed
saddle-point methods. The semiclassical action has been modified in order to
account for four different physical processes. In two of them, the first
electron is released and rescatters off the same center $C_{j}(j=1,2)$. In
the remaining processes, it leaves from a center $C_{j}$ and rescatters with
a different center $C_{\nu },$ $\nu \neq j.$

We placed particular emphasis on the quantum interference between such
processes, and on how it affects the differential electron momentum
distributions, as functions of the momentum components $p_{(n)||}(n=1,2)$
parallel to the laser-field polarization. For that purpose, we considered
parallel-aligned molecules, for which the interference patterns are sharpest
\cite{FSLY2008}. We found that the contributions of the processes in which
the first electron starts at the same center $C_{j}$, regardless of whether
it returns to the site of its release or to the other center $C_{\nu }$,
yield interference patterns which fulfill the overall interference condition
(\ref{fringespar}). There exist, however, differences for both velocity and
length-gauge formulations of the SFA. Indeed, while in the velocity gauge
the contributions of the above-stated orbits leads to the same distributions
as in the overall case, this affirmative cannot be made in the length-gauge
situation. In this latter case, there are additional potential energy
shifts, which sink or increase the potential barrier through which the first
electron tunnels. This favors the contributions of a particular center in
the molecule, and suppresses those from the other center. As the electric
field changes sign, so do the shifts. This leads to asymmetries in the
electron-momentum distributions.

The quantum interference between other types of processes leads either to
quite different interference patterns, or to momentum distributions that
strongly resemble those obtained for a single atom. In particular, a
resemblance to single-atom distributions is observed if only the processes
for which the first electron rescatters at the same center $C_{j}$ are
considered. The fact that interference fringes are absent is in agreement
with the double-slit physical picture, as there is no electron emission at
spatially separated centers for such processes.

Sharp interference fringes, on the other hand, are observed if we consider
topologically similar scattering scenarios, involving either one or two
centers, and the velocity gauge. This is expected since, in this case, the
first electron rescatters at spatially separated centers. It is worth
noticing, however, that the interference patterns, and thus the
electron-momentum distributions, look quite different from those obtained if
all contributions are taken. Physically, this indicates that we can not
single out the interference between such processes as being the most
relevant. In the length gauge, in principle, there are also interference
patterns. In practice, however, we only see slight distortions, as compared
to the single-atom case. This is due to the additional potential-energy
shifts, which sink or increase the potential barrier through which the first
electron must tunnel in order to reach the continuum. Hence, the
contributions starting at a center $C_{j}$ in which the barrier has been
sunk will be far more prominent than those from a center $C_{\nu }$ in which
it has been raised.

The above-stated issues contribute to the present controversy involving the
above-mentioned potential-energy shifts \cite%
{PRACL2006,F2007,DM2008,DM2006,dressedSFA,dressedSFA2,BCM2007}. Indeed, it
is not clear whether they are only an artifact of the strong-field
approximation, or whether they possess some physical meaning. For instance,
in \cite{F2007,dressedSFA2} it has been shown that, if such shifts are not
present, there exists a breakdown in the double-slit interference patterns
for high-order harmonic generation. Specifically in \cite{dressedSFA2}, it
was argued that, for small internuclear distances, there is no physical
justification for dressing the electronic bound states with a linear phase
factor. In fact, according to \cite{dressedSFA2} a quadratic phase factor
would be the correct form of dressing, and would physically correspond to
the quadratic Stark effect. On the other hand, in \cite{PRACL2006}, it was
shown that, for internuclear distances of the order of the electron
excursion amplitude, the additional potential-energy shifts led to an
unphysical increase in the high-harmonic intensities and cutoff energies.
Furthermore, in \cite{BCM2007} it was demonstrated that a linear dressing in
the electronic bound states is necessary in order cancel out the extra
potential-energy shifts and to restore, among others, translation
invariance. In this specific work, \ one argument against the existence of
the potential-energy shifts would be that, for a homonuclear molecule, one
would intuitively expect the symmetries $|M_{11}+M_{21}|^{2}=$ $%
|M_{22}+M_{12}|^{2}$ and $|M_{11}+M_{12}|^{2}=$ $|M_{22}+M_{21}|^{2}$ to
hold. Due to the fact that, in the length gauge, the imaginary part of $%
t^{\prime }$ is shifted in different ways for different centers in molecule,
we have seen that, instead, the distributions starting at a specific center $%
C_{j}$ are a reflection upon $(p_{1||},p_{2||})\rightarrow $ $%
(-p_{1||},-p_{2||})$ of those starting at the other center $C_{\nu }.$

Finally, we would like to comment on the similarities and differences
between the present results and those reported in our previous work \cite%
{F2007,F2008}, on the double-slit quantum interference for HHG in diatomic
molecules. Therein, even though there was strong evidence that the
interference maxima and minima in the spectra were due to the transition
probability $|M_{jj}+M_{j\nu }|^{2},$ related to the interference of the
orbits starting at the same center and finishing at different centers, there
were several ambiguities. Firstly, this information could only be extracted
in the length gauge, due to limitations of the SFA. Within this approach,
there is a breakdown in the interference patterns if the velocity gauge is
employed to compute the high-harmonic yield. Furthermore, similarly to the
present case, the length-gauge formulation of the SFA exhibits additional
potential-energy shifts which make the ionization probability far more
probable from one center than from the other. This holds even for very small
internuclear distances \cite{F2008}, and could be the reason why the
interference patterns would not be present also in the transition
probabilities $|M_{jj}+M_{\nu \nu }|^{2},$ or $|M_{j\nu }+M_{\nu j}|^{2},$
involving one or two-center orbits, respectively. By providing an additional
pathway for the electron to reach the continuum, using attosecond pulses, we
could verify that these processes led to maxima and minima if the
potential-energy shifts could be overcome. Since, however, we were modifying
the physics of the problem, we could not reach a definitive conclusion.

Fortunately, many of these ambiguities are not present for laser-induced
nonsequential double ionization. This is possibly due to the fact that this
phenomenon intrinsically involves electron-electron correlation, whereas
high-order harmonic generation may be modeled within the single-active
electron approximation. Indeed, although the distributions computed in this
work employing the topologically similar scattering scenarios may, in some
cases, exhibit interference patterns, these do not agree with the overall
interference conditions. Furthermore, in contrast to the high-order harmonic
case, there is no breakdown of the interference patterns in the
velocity-gauge situation. This allows one to assess the influence of the
topologically similar scenarios on the NSDI electron momentum distributions
in the absence of the above-mentioned potential energy shifts. Finally, it
is rather interesting that, in the length gauge, the influence of the
potential energy shifts can be directly mapped into the electron momentum
distributions for the processes starting at the same center. For instance,
if the electron started at a center $C_{1}$, for the parameters used in this
work, there is a suppression of the yield in the positive momentum region
and an enhancement for positive parallel momentum, whereas if it has started
at $C_{2}$ the situation will be reversed, i.e., there will be an
enhancement in the region of positive parallel momenta and a suppression for
$p_{(n)||}<0$ $(n=1,2)$. This means that different potential energy shifts
lead to an enhancement or a suppression of the distributions in distinct
momentum regions.

Hence, if these processes could be isolated in a realistic scenario, one
could in principle determine from which side of the molecule the first
electron reached the continuum. Furthermore, if no asymmetry is observed,
the existence of the potential-energy shifts and the validity of the
length-gauge SFA electron momentum distributions could be ruled out in the
context of diatomic molecules. For that purpose, it would be necessary to
suppress all the the orbits starting from one center of the molecule in the
overall distributions . This could be achieved if, for instance, the system
is prepared in such a way that the initial state of the first electron is
localized in one center of the molecule, or if the external laser field is
taken as a few-cycle pulse with an adequate choice of carrier-envelope phase.

\ack We are grateful to T. Shaaran, C. Ruiz, X. Liu and W. Yang for useful
discussions. This work has been financed by the UK EPSRC (Advanced
Fellowship, Grant no. EP/D07309X/1).


\section*{References}

\begin{thebibliography}{99}
\bibitem{attomol} Niikura H, L\'{e}gar\'{e} F, Hasbani R, Bandrauk A D,
Ivanov M Yu, Villeneuve D M and Corkum P B 2002 \textit{Nature }\textbf{417}
917

\hspace*{-0.5cm} Niikura H, L\'{e}gar\'{e} F, Hasbani R, Ivanov M Yu,
Villeneuve D M and Corkum P B, 2003 \textit{Nature} \textbf{421} 826

\hspace*{-0.5cm} Itatani J, Levesque J, Zeidler D, Niikura H, P\'{e}pin H,
Kieffer J C, Corkum P B and Villeneuve D M, 2004 \textit{Nature }\textbf{432}
867

\hspace*{-0.5cm} Baker S, Robinson J S, Haworth C A, Teng H, Smith R A,
Chiril\u{a} C C, Lein M, Tisch J W G, Marangos J P, 2006 \textit{Science}
\textbf{312} 424

\bibitem{tstep} Corkum P B, 1993 \textit{Phys. Rev. Lett.} \textbf{71} 1994

\hspace*{-0.5cm} Kulander K C, Schafer K J, and Krause J L in: B. Piraux et
al. eds., \emph{Proceedings of the SILAP conference}, (Plenum, New York,
1993)

\bibitem{Scrinzi2006} Scrinzi A, Ivanov M Yu, Kienberger R, and Villeneuve D
M, 2006 \textit{J. Phys. B} \textbf{39} R1

\bibitem{doubleslit} Lein M, Hay N, Velotta R, Marangos J P, and Knight P L,
2002 \textit{Phys. Rev. Lett.} \textbf{88} 183903

\hspace*{-0.5cm} Lein M, Hay N, Velotta R, Marangos J P, and Knight P L 2002
\textit{Phys. Rev. A} \textbf{66} 023805

\hspace*{-0.5cm} Spanner M, Smirnova O, Corkum P B and Ivanov M Yu, 2004
\textit{J. Phys. B}\textbf{\ 37} L243

\bibitem{KBK98} Kopold R, Becker W and Kleber M, 1998 \textit{Phys. Rev. A}
\textbf{58} 4022

\bibitem{Usach2006} Usachenko V I, Pyak P E, and Chu Shih-I, 2006 \textit{%
Laser Phys.} \textbf{16} 1326

\bibitem{PRACL2006} Chiril\u{a} C C and Lein M, 2006 \textit{Phys. Rev. A}
\textbf{73} 023410

\bibitem{HBF2007} Hetzheim H, Figueira de Morisson Faria C, and Becker W
2007 \textit{Phys. Rev. A} \textbf{76} 023418

\bibitem{F2007} Figueira de Morisson Faria C, 2007 \emph{Phys. Rev. A}
\textbf{76} 043407

\bibitem{F2008} Figueira de Morisson Faria C, 2007, arXiv: 0810.5251

\bibitem{DM2008} Busulad\v{z}i\'{c} M, Gazibegovi\'{c}-Busulad\v{z}i\'{c} A,
Milo\v{s}evi\'{c} D B, and Becker W 2008 \textit{Phys. Rev. Lett.} \textbf{%
100} 203003

\bibitem{SFAold} Lewenstein M, Balcou Ph, Ivanov M Yu, L'Huillier A and
Corkum P B, 1994 \textit{Phys. Rev. A} \textbf{49} 2117

\hspace*{-0.5cm} Becker W, Long S, and McIver J K 1990 \textit{Phys. Rev. A}
\textbf{41} 4112

\hspace*{-0.5cm} Becker W, Long S, and McIver J K 1994 \textit{Phys. Rev. A}
\textbf{50} 1540

\hspace*{-0.5cm} Lewenstein M, Kulander K C, Schafer K J and Bucksbaum P
1995 \textit{Phys. Rev. A} \textbf{51} 1495

\bibitem{Usachenko} Usachenko V I, and Chu S I 2005 \textit{Phys. Rev. A}
\textbf{71} 063410

\bibitem{moreCL} Lein M 2005 \textit{Phys. Rev. Lett.} \textbf{94} 053004

\hspace*{-0.5cm} Chiril\u{a} C C and Lein M 2006 \textit{J. Phys. B} \textbf{%
39} S437

\hspace*{-0.5cm} Chiril\u{a} C C and Lein M, 2007 \textit{J. Mod. Opt.}
\textbf{54} 1039

\bibitem{MBBF00} Muth-B\"{o}hm J, Becker A, and Faisal F H M, 2000 \textit{%
Phys. Rev. Lett.} \textbf{85} 2280

\hspace*{-0.5cm} Jar\'{o}n-Becker A, Becker A, and Faisal F H M 2004 \textit{%
Phys. Rev. A} \textbf{69} 023410

\hspace*{-0.5cm} Requate A, Becker A and Faisal F H M 2006 \textit{Phys.
Rev. A} \textbf{73} 033406

\bibitem{Madsen} Kjeldsen T K and Madsen L B 2004 \textit{J. Phys. B}
\textbf{37} 2033

\hspace*{-0.5cm} Kjeldsen T K and Madsen L B 2005 \textit{Phys. Rev. A}
\textbf{71} 023411

\hspace*{-0.5cm} Kjeldsen T K and Madsen L B 2005 \textit{Phys. Rev. Lett}.
\textbf{95} 073004

\hspace*{-0.5cm} Kjeldsen T K, Bisgaard C Z, Madsen L B, Stapelfeld H 2005
\textit{Phys. Rev. A} \textbf{71} 013418

\hspace*{-0.5cm} Madsen C B and Madsen L B 2006 \textit{Phys. Rev. A}
\textbf{74} 023403

\bibitem{Kansas} Zhou X, Tong X M, Zhao Z X and Lin C D 2005 \textit{Phys.
Rev. A }\textbf{71} 061801(R)

\hspace*{-0.5cm} Zhou X, Tong X M, Zhao Z X and Lin C D2005\textit{\ Phys.
Rev. A} \textbf{72} 033412

\bibitem{DM2006} Milo\v{s}evi\'{c} D B, 2006 \textit{Phys. Rev. A }\textbf{74%
} 063404

\bibitem{FSLY2008} Figueira de Morisson Faria C, Shaaran T, Liu X and Yang
W, arXiv:0806.4856 [atom.ph]

\bibitem{NSDIsymm} Eremina E, Liu X, Rottke H, Sandner W, Sch\"{a}tzel M G,
Dreischuch A, Paulus G G, Walther H, Moshammer R and Ullrich J 2004 \textit{%
Phys. Rev. Lett.} \textbf{92} 173001

\bibitem{NSDIalign} Zeidler D, Staudte A, Bardon A B, Villeneuve D M, D\"{o}%
rner R, and Corkum P B 2005 \textit{Phys. Rev. Lett.} \textbf{95} 203003

\bibitem{FKS96} Fring A, Kostrykin V and Schrader R 1996 \textit{J. Phys. B.
}\textbf{29} 5651

\hspace*{-0.5cm} Bauer D, Milo\v{s}evi\'{c} D B and Becker W 2005 \textit{%
Phys. Rev. A} \textbf{72} 023415

\bibitem{atiuni} Figueira de Morisson Faria C, Schomerus H and Becker W 2002
\textit{Phys. Rev. A} \textbf{66} 043413

\bibitem{FSLB2004} Figueira de Morisson Faria C, Schomerus H, Liu X, and
Becker W 2004 \textit{Phys. Rev. A} \textbf{69} 043405

\bibitem{FL2005} Figueira de Morisson Faria C, and Lewenstein M 2005 \textit{%
J. Phys. B} \textbf{38} 3251

\bibitem{dressedSFA} Smirnova O, Spanner M and Ivanov M 2006 \textit{J.
Phys. B} \textbf{39} S307

\bibitem{dressedSFA2} Smirnova O, Spanner M and Ivanov M 2007 \textit{J.
Mod. Opt.}\textbf{54} 1019

\bibitem{BCM2007} Becker W, Chen J, Chen S G, and Milo\v{s}evi\'{c} D B,
2007 \textit{Phys. Rev. A} \textbf{76}, 033403

\bibitem{LC2008} Chipperfield L, 2008, Ph.D. thesis, Imperial College London
\end{thebibliography}
\end{document}